\documentclass{elsart}

 \usepackage{epsfig}

\usepackage{amssymb}

\begin{document}

\begin{frontmatter}



\title{Electromagnetic mass model admitting conformal motion}

\author[label1]{Saibal Ray}\footnote{Corresponding author (E-mail: saibal@iucaa.ernet.in).},
\author[label2]{A A Usmani},
\author[label3]{F Rahaman},
\author[label4]{M Kalam},
\author[label5]{K Chakraborty},

\address[label1]{Department of Physics, Barasat Government College,
Barasat 700124, North 24 Parganas, West Bengal, India}
\address[label2]{Department of Physics, Aligarh Muslim University, Aligarh 202 002,
Uttar Pradesh, India}
\address[label3]{Department of Mathematics, Jadavpur University, Kolkata 700 032,
 West Bengal, India}
\address[label4]{Department of Physics, Netaji Nagar College for Women, Regent
Estate, Kolkata 700 092, West Bengal, India}
\address[label5]{Department of Mathematics, Jadavpur University, Kolkata 700 032,
 West Bengal, India}

\begin{abstract}
We study charged fluid spheres under the $4$-dimensional
Einstein-Maxwell space-time. The solutions thus obtained admitting
conformal motion. We also investigate whether the solutions set
provide electromagnetic mass models such that the physical
parameters including the gravitational mass arise from the
electromagnetic field alone. In this connection three cases are
studied here in detail with the propositions: (1) $p = - \rho$,
(2) $\sigma e^{\lambda/2} = \sigma_0$ and (3) $8 \pi p - E^2 =
p_0$ where $\rho$, $p$, $\sigma$ are respectively the usual matter
density, fluid pressure and charge density of the spherical
distribution. Based on these assumptions several features are
explored which seems physically very interesting.
\end{abstract}

\end{frontmatter}

\section{Introduction}
There is a fairly long history of investigations regarding the
nature of the mass of electron. While studying the interaction of
charged particles Thomson~\cite{Thomson1881} found that the
kinetic energy of a charged sphere increases by its motion through
a medium of finite specific inductive capacity. He pointed out
that the increase in the kinetic energy was due to the self
induced magnetic field of the charged sphere and came to the
conclusion that ``the effect of electrification is the same as if
the mass of the sphere were increased ...". This investigation was
improved upon by Heaviside~\cite{Heaviside1889} showing that the
mass of a uniformly moving charged body varied with velocity.
Searle~\cite{Searle1897} extended this again as that the energy of
a charged body and hence its mass increases with velocity. Later
on
Kaufmann~\cite{Kaufmann1901a,Kaufmann1901b,Kaufmann1901c,Kaufmann1902a,Kaufmann1902b}
conducted a series of experiments on beta rays and established the
dependence of the electron mass on velocity.
He~\cite{Kaufmann1901b} showed that one-third of the fast moving
electron mass is of electromagnetic origin. However,
Abraham~\cite{Abraham1902} speculated that the electron mass was
completely of electromagnetic origin. This result insisted
Kaufmann~\cite{Kaufmann1902b} to re-analyze his experimental data
and he found that the mass of the electron is purely of
electromagnetic origin.

Under this illuminated background, therefore,
Lorentz~\cite{Lorentz1904} proposed his model for extended
electron and conjectured that ``there is no other, no `true' or
`material' mass,'' and thus provides only `electromagnetic masses
of the electron'\footnote{It would be historically very
interesting to note that even after these profound theoretical and
experimental results Einstein~\cite{Einstein1919} himself believed
that ``... of the energy constituting matter three-quarters is to
be ascribed to the electromagnetic field, and one-quarter to the
gravitational field''.}. Wheeler~\cite{Wheeler1962} and
Wilczek~\cite{Wilczek1999} pointed out that electron has a ``mass
without mass'' whereas Feynman, Leighton and
Sands~\cite{Feynman1964} termed this type of models as
``electromagnetic mass models''. Following the idea of {\it
electromagnetic mass} (EMM) a lot of works have been carried out
by several
authors~\cite{Florides1962,Cooperstock1978,Tiwari1984,Gautreau1985,Gron1985,Leon1987,Tiwari1991,Ray2004a,Ray2006}
under the framework of general relativity where space-time
geometry is assumed to be associated with the presence of matter.

On the other hand to search the natural relation between geometry
and matter through the Einstein equations it is convenient to use
inheritance symmetry. The well known inheritance symmetry is the
symmetry under conformal killing vectors (CKV) which can be given
by $L_\xi g_{ij} = \psi g_{ij}$ where $L$ is the Lie derivative
operator and $\psi$ is the conformal factor. Here it is supposed
that the vector $\xi$ generates the conformal symmetry and then
the metric $g$ is conformally mapped onto itself along $\xi$. In
this connection it is to be noted that neither $\xi$ nor $\psi$
need to be static even though one considers a static
metric~\cite{Bohmer2008a,Bohmer2008b}. It can be seen in the
literature that due to this and several other properties CKVs
provide a deeper insight into the space-time geometry connected to
astrophysical and cosmological
realm~\cite{Maartens1996,Mars2002,Mak2004a,Harko2004,Mak2004b,Neves2006}.

Now the above works on EMM
models~\cite{Florides1962,Cooperstock1978,Tiwari1984,Gautreau1985,Gron1985,Leon1987,Tiwari1991,Ray2004a,Ray2006}
have been performed either in $4D$ or higher dimensional
space-time under general relativity. Therefore, our motivation in
the present investigation is to include CKV and see whether
conformal motion admits EMM models or not. To do so we have
considered here static spherically symmetric charged perfect fluid
distribution under $4D$ general relativity and studied three
cases: (1) $p = - \rho$, (2) $\sigma e^{\lambda/2} = \sigma_0$ and
(3) $8 \pi p - E^2 = p_0$ where $\rho$, $p$, $\sigma$ are
respectively the usual matter density, fluid pressure and charge
density of the spherical distribution. The scheme of the present
investigations are follows:  Sections 2 and 3 are related to the
Einstein field equations with CKV and their solutions for the
three cases. In the concluding Section 4, some remarks are made.

\section{Einstein-Maxwell field equations in $4$-dimensional space-time}
The static spherically symmetric space-time is taken as
\begin{equation}
               ds^2=  - e^{\nu(r)} dt^2+ e^{\lambda(r)} dr^2+r^2( d\theta^2+sin^2\theta
               d\phi^2)
\end{equation}
where $\nu$ and $\lambda$ are the metric potentials and functions
of radial coordinate $r$ only.

Now, the Einstein field equations for the case of charged fluid
source are
\begin{equation}
{G^{i}}_{j} = {R^{i}}_{j} - \frac{1}{2}{{g^{i}}_{j}} R = -\kappa
\left[{{T^{i}}_{j}}^{(m)} + {{T^{i}}_{j}}^{(em)}\right]
\end{equation}
where ${T^{i}}_{j}^{(m)}$ and ${T^{i}}_{j}^{(em)}$ are,
respectively, the energy-momentum tensor components for the matter
source and electromagnetic field. The explicit forms of these
tensors are given by
\begin{equation}
{{T^{i}}_{j}}^{(m)} = (\rho + p) u^{i}u_{j} + p{g^{i}}_{j},
\end{equation}
\begin{equation}
{{T^{i}}_{j}}^{(em)} = -\frac{1}{4\pi}\left[ F_{jk}F^{ik} -
\frac{1}{4} {g^{i}}_{j}F_{kl} F^{kl}\right]
\end{equation}
where $\rho$, $p$ and $u^{i}$ are, respectively, matter-energy
density, fluid pressure and velocity four-vector of a fluid
element (with $u_{i}u^{i}=1$).

The Maxwell electromagnetic field equations are given by
\begin{equation}
{[{(- g)}^{1/2} F^{ij}],}_{j} = 4\pi J^{i}{(- g)}^{1/2},
\end{equation}
\begin{equation}
 \label{eqx} F_{[ij,k]} = 0,
\end{equation}
where the electromagnetic field tensor $F_{ij}$ is related to the
electromagnetic potentials through the relation $ F_{ij} = A_{i,j}
- A_{j,i} $ which is equivalent to the equation~(\ref{eqx}). In
the above equations $J^{i}$ is the current four-vector satisfying
$J^{i} = \sigma u^{i}$, where $\sigma$ is the charge density, and
$ \kappa = 8 \pi $. We have considered the relativistic unit for
which $G = c = 1$. Here and in what follows a comma denotes the
partial derivative with respect to the coordinates.

In view of above, therefore, the Einstein-Maxwell field equations
can be given as
\begin{equation}\label{eqE1} e^{-\lambda}
\left[\frac{\lambda^\prime}{r} -
\frac{1}{r^2}\right]+\frac{1}{r^2}= 8\pi \rho + E^2,
\end{equation}
\begin{equation}
\label{eqE2} e^{-\lambda}
\left[\frac{1}{r^2}+\frac{\nu^\prime}{r}\right]-\frac{1}{r^2}=
8\pi p - E^2,
\end{equation}
\begin{equation}
\label{eqE3} \frac{1}{2} e^{-\lambda}
\left[\frac{1}{2}(\nu^\prime)^2+ \nu^{\prime\prime}
-\frac{1}{2}\lambda^\prime\nu^\prime + \frac{1}{r}({\nu^\prime-
\lambda^\prime})\right] =8\pi p + E^2
\end{equation}
and
\begin{equation}
\label{eqMx}(r^2E)^\prime = 4\pi r^2 \sigma e^{\lambda/2}.
\end{equation}
The electric field $E$ in the above equation~(\ref{eqMx}) can
equivalently be expressed in the form
\begin{equation}
\label{eqMy}E(r) = \frac{1}{r^2}\int_0^r 4\pi r^2 \sigma
e^{\frac{\lambda}{2}}dr = \frac{q(r)}{r^2}
\end{equation}
where $q(r)$ is the total charge of the sphere under
consideration.

It is interesting to note that the equations (7) and (8) provide
an essential relationship between the metric potentials and the
physical parameters $\rho$ and $p$ as follows
\begin{equation}
e^{-\lambda}(\nu^{\prime} + \lambda^{\prime}) = 8 \pi r( \rho +
p).
\end{equation}
 Again, equation (7) may be expressed in the general form as
\begin{equation}
e^{-\lambda} = 1 -  \frac{2M(r)}{r},
\end{equation}
where M(r), known as the active gravitational mass, takes the form
\begin{equation}
\label{eqM}M(r) = 4 \pi \int_{0}^{r}\left[ \rho +  \frac{E^2}{8
\pi}\right] r^2 dr.
\end{equation}

Let us now consider the problem of charged fluid sphere under
conformal motion through CKV which can be given by
\begin{equation}
\label{eqckv}
L_\xi g_{ij} =\xi_{i;j}+ \xi_{j;i} = \psi g_{ij}.
\end{equation}
The above equations~(\ref{eqckv}) give the following set of
expressions

$\xi^1 \nu^\prime =\psi$,

$\xi^4  = C_1 = constant$,

$\xi^1  = \frac{\psi r}{2}$,

$\xi^1 \lambda ^\prime + 2 \xi^1 _{,1}   =\psi$.

These, therefore, readily imply that
\begin{equation}
\label{eqK1}
e^\nu  =C_2^2 r^2,
\end{equation}
\begin{equation}
\label{eqK2}
e^\lambda  = \frac{C_3^2}{{\psi}^2},
\end{equation}
\begin{equation}
\xi^i = C_1 \delta_4^i + \left(\frac{\psi r}{2}\right)\delta_1^i
\end{equation}
where $C_2$ and $C_3$ are integration constants.

\section{Electromagnetic mass models with conformal motions}
 Now using solutions~(\ref{eqK1}) and~(\ref{eqK2}), the equations~(\ref{eqE1}) - (\ref{eqE3}) take the
following form
\begin{equation}
\label{eqx1}
\frac{1}{r^2}\left[1 - \frac{\psi^2}{C_3^2}
\right]-\frac{2\psi\psi^\prime}{rC_3^2}= 8\pi \rho + E^2,
\end{equation}
\begin{equation}
\label{eqx2}
\frac{1}{r^2}\left[1 - \frac{3\psi^2}{C_3^2} \right]=
- 8\pi p + E^2,
\end{equation}
\begin{equation}
\label{eqx3}
\frac{\psi^2}{C_3^2r^2}+\frac{2\psi\psi^\prime}{rC_3^2} =8\pi p +
E^2.
\end{equation}

From the above equations~(\ref{eqx1}) - (\ref{eqx3}), in a
straight forward way, one can get the values for $E$, $\rho$ and
$p$ as
\begin{equation}
\label{eqy1}
E^2 = \frac{1}{2} \left[ \frac{1}{r^2}\left(1 -
\frac{2\psi^2}{C_3^2}
\right)+\frac{2\psi\psi^\prime}{rC_3^2}\right],
\end{equation}
\begin{equation}
\label{eqy2}
8\pi \rho  =  \frac{1}{2r^2}
-\frac{3\psi\psi^\prime}{rC_3^2},
\end{equation}
\begin{equation}
\label{eqy3}
8\pi p = -\frac{1}{2r^2} +
\frac{\psi\psi^\prime}{rC_3^2} +\frac{2\psi^2}{r^2C_3^2}.
\end{equation}

In the following subsections we shall consider three different
cases to get exact analytical solutions in connection to
relativistic charged fluid spheres with conformal motion.

\subsection{Case-I : $p = - \rho$}
It is already mentioned in the introductory part that there is a
special kind of solution known in the literature as {\it
electromagnetic mass} (EMM) models where all the physical
parameters, including the gravitational mass, are arising from the
electromagnetic field alone have been extensively studied by
several
researchers~\cite{Lorentz1904,Feynman1964,Florides1962,Cooperstock1978,Tiwari1984,Gautreau1985,Gron1985,Leon1987,Tiwari1991,Ray2004a,Ray2006}.
In this connection it is interesting to note that most of these
investigators exploit an equation of state $p = -\rho$ with the
equation of state parameter $\omega=-1$ which is a very common
feature in the context of accelerating phase of the present
Universe \footnote{In general, the matter density $\rho>0$ and
fluid pressure $p<0$. However, there are some special cases
available in the literature where $\rho<0$ and hence
$p>0$~\cite{Cooperstock1989,Bonnor1989,Herrera1994,Ray2004b}.}.
The equation of state of this type implies that the matter
distribution under consideration is in tension and hence the
matter is known in the literature as a `false vacuum' or
`degenerate vacuum' or
`$\rho$-vacuum'~\cite{Davies1984,Blome1984,Hogan1984,Kaiser1984}.

To consider the above mentioned equation of state let us use the
equations~(\ref{eqx1}) and (\ref{eqx2}) which provide the unique
relation
\begin{equation}
\label{key1} \frac{2\psi}{r^2C_3^2}(\psi-r\psi^\prime) = 8 \pi r(
\rho + p).
\end{equation}
Therefore, the equation~(\ref{key1}), due to the {\it ansatz}
$\rho + p = 0$, gives the value for $\psi$ as either $\psi =0$ or
$\psi = \psi_0 r$, where $\psi_0$ is an integration constant.

Now, by the use of non-zero value of $\psi$ the exact analytical
form for all the physical parameters can be given as
\begin{equation}
8\pi \rho  =  \frac{1}{2r^2} - 3C_4^2,
\end{equation}
\begin{equation}
8\pi p  =  -\frac{1}{2r^2} + 3C_4^2,
\end{equation}
\begin{equation}
\label{eqE1} E^2 = \frac{q^2}{r^4} = \frac{1}{2r^2},
\end{equation}
\begin{equation}
\label{eqs}\sigma  = \frac{C_4}{4 \pi \sqrt{2} r},
\end{equation}
\begin{equation}
e^\nu  = e^{-\lambda}= C_4^2r^2,
\end{equation}
where $\psi_0/C_3=C_4$.

\begin{figure}
\begin{center}
\vspace{0.5cm} \psfig{file=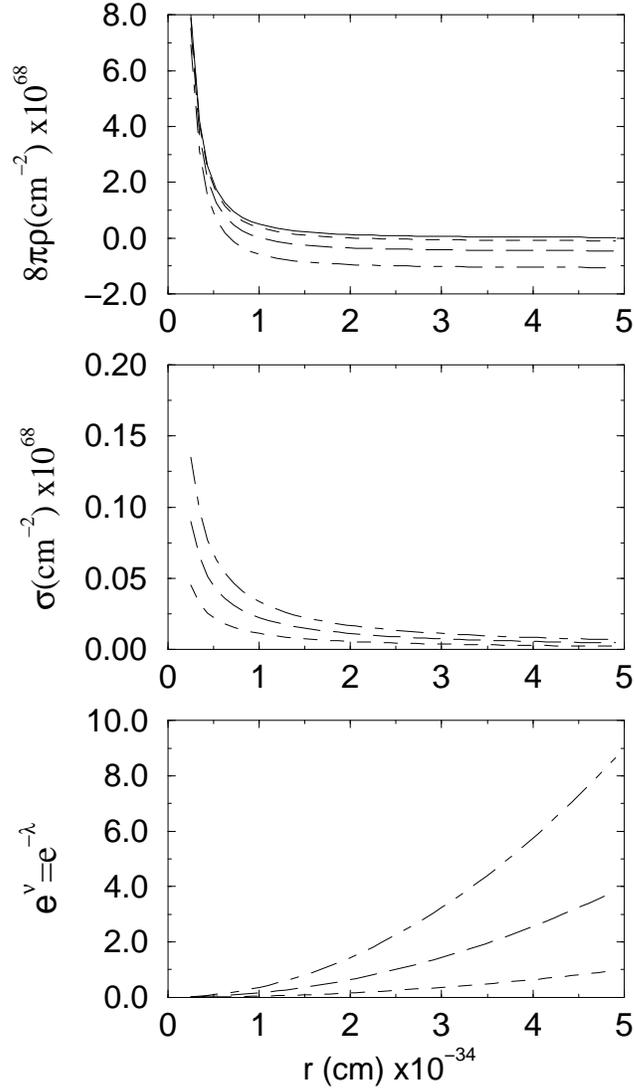,width=0.6\textwidth}
\caption{The upper, middle and lower panels represent plots for
$8\pi\rho~(=-8\pi p)$, $\sigma$ and $e^\nu~(=e^{-\lambda})$.
 The solid, dashed, long-dashed and chain curves correspond to
$C_4= 0.0$, $0.2$, $0.4$ and $0.6$ cm$^{-1}$, respectively. }
\label{fig:1}
\end{center}
\end{figure}

 Hence, the total gravitational mass $m(r=a)$, which we get after
 matching of the interior solution to
the exterior Reissner-Nordstr\"{o}m solution on the boundary, can
be calculated as
\begin{equation}
\label{eqm1}
m(a) =M(a) + \frac{q(a)^2}{2a} = \frac{1}{2\sqrt2}(3
- 4C_4^2q^2)q.
\end{equation}
It can be observed from the equation~(\ref{eqs}) that vanishing
charge density $\sigma$ makes $\psi_0=0=C_4$. This implies
$E^2=8\pi\rho=-8\pi p=1/2r^2$ and $e^{\nu}=e^{-\lambda}=0$. Thus
$\rho$, $p$ and $E^2$ have $1/r^2$ behaviour with proportionality
constants $1/16\pi$, $-1/16\pi$ and $1/2$, respectively. In case
of non-zero $C_4$ the values of $8\pi\rho$ and $8\pi p$ would be
shifted in opposite directions by an amount $3C^2_4$ with respect
to $E^2$. We also observe that $\sigma$ has got $1/r$ behaviour
and $e^{\nu}=e^{-\lambda}\propto r^2$. Interestingly, the
vanishing charge $q$ does turn the total gravitational mass $m$,
as expressed in the above equation~(\ref{eqm1}), into an {\it EMM}
as expected under the ansatz $\rho+p=0$. In a similar way all the
other physical parameters involved in the equations (26) - (30)
vanish due to vanishing charge. These results imply that conformal
motion does admit this type of EMM model. However, a close
observation demands that as mass must not be negative so the above
expression for mass puts a limit on $C_4$ with a fixed charge $q$,
as the second term must be smaller than the first one, which
implies $C_4<\sqrt3/2q$. Thus charge $q$ limits the
proportionality constants for $\sigma$, $e^{\nu}$ and
$e^{-\lambda}$ and also the departure of $8\pi\rho$ and $8\pi p$
from $E^2$. We plot all these observable parameters in figure
\ref{fig:1}.

\subsection{Case-II: $\sigma e^{\lambda/2} = \sigma_0 $}
Let us consider here the {\it ansatz} in such a way that the
charge density remains constant, say
$\sigma_0$~\cite{Tiwari1984,Ray2004b,Ray2006}. This is the charge
density of the spherical fluid distribution at the centre $r=0$.

For the above assumption of constant charge density case the
equation~(\ref{eqMy}) leads to a proportionality  between $E^2$ and $r^2$
as given by
\begin{equation}
\label{eqE2}
E^2 =  A r^{2},
\end{equation}
where $A=16 \pi^2 \sigma_0^2 /9$ is the proportionality constant.

Substitution of equation~(\ref{eqy1}) in
equation~(\ref{eqE2}) yields
\begin{equation}
\label{eqy12}
2A r^{2} =
\left[ \frac{1}{r^2}\left(1 - \frac{2\psi^2}{C_3^2}
\right)+\frac{2\psi\psi^\prime}{rC_3^2}\right].
\end{equation}

By solving above equation~(\ref{eqy12}) we get
\begin{equation}
 \psi ^2  =  Cr^2 +\frac{C_3^2}{2} \left[1+ 2A r^{4
 }\right]
\end{equation}
where $C$ is an integration constant.

Expressions for $\rho$ and $p$ in equations~(\ref{eqy2})
and~(\ref{eqy3}), thus take the following forms
\begin{equation}
\label{eqrho2} 8\pi \rho  =  \frac{1}{2r^2} - 2A r^{2} -3C_5,
\end{equation}
\begin{equation}
\label{eqp2} 8\pi p  = \frac{1}{2r^2}+ 4A r^{2}+3C_5
\end{equation}
where $C_5=C/C_3^2$.

Therefore, the total gravitational mass $m(r=a)$ here can be given
by
\begin{equation}
\label{eqm2} m(a) = \frac{a}{4} - \frac{C_5 a^3}{2} +
\frac{12A}{5}a^{5}.
\end{equation}
The condition, $m(a)>0$ restricts $C_5$ as
\begin{equation}
C_5 <\frac{1}{2a^2} + \frac{24A}{5}a^{2}.
\end{equation}
For the typical values of the physical parameters $q=1.38 \times
10^{-34}$ cm, $a=1.0 \times 10^{-16}$ cm and $\sigma_0=3q/4\pi
a^3=3.29 \times 10^{13}$ cm$^{-2}$ (in relativistic units) we get
$A=1.9\times 10^{28}$cm$^{-2}$ and hence $C_5 <5.0\times 10^{31}$.
The equation~(\ref{eqrho2}) for $8\pi\rho$ and
equation~(\ref{eqp2}) for $8\pi p$ are represented by the
interference of three terms. The first and second term become
comparable to each other around $r=r_{critical}$. For $r
<<r_{critical} $, second term in both the equations~(\ref{eqrho2})
and (\ref{eqp2}) are negligibly small compared to the first term
i.e. $1/r^2$. In this case, equation~(\ref{eqrho2}) reads as
$8\pi\rho= 1/r^2 -3C_5$, which is the same expression obtained for
Case-I ($p=-\rho$) but with a very small limit on the integration
constant. On the other hand, equation~(\ref{eqp2}) reads as $8\pi
p= 1/r^2 +3C_5$ but with a different sign for the integration
constant. Here, unlike Case-I, we do not have the condition
$p=-\rho$. For the case, $r>>r_{critical}$, it is the second term
that dominates and the first term becomes negligibly small.

Now, the relationship between the electric charge and intensity of
electric field being $q(r)^2=E^2r^4$, the equation~(\ref{eqE2})
reduces for the total charge to
\begin{equation}
\label{eqq2}
q(a) = \sqrt{A} a^{3}.
\end{equation}
Here $q=0$ does not imply $a=0$ as we obtained it in the previous
case ($q=a/\sqrt {2}$ in the equation (28)) rather it provides the
result $\sigma_0=0$. Therefore, substitution of this
$a\approx([1/\sqrt{A}]^{1/3}=[3q/4 \pi \sigma_0]^{1/3})$ in the
equation~(\ref{eqm2}) does not make the above gravitational mass
to vanish for the vanishing electric charge. However, for
$\sigma_0=0$ the constant $A$ becomes zero. This immediately
implies that for the gravitational mass to vanish the condition to
be imposed here is $C_5=1/2a^2$ whereas for the energy density and
fluid pressure this is $C_5=\pm 1/6r^2$. Therefore, we get a
unique condition for {\it EMM} model which can be given by
$r_{critical}=a/\sqrt3$ and can be referred as the {\it critical
radius} for EMM in connection to Lorentz's type extended electron.
This means that not the mass of the whole spherical body rather
only a sphere of radius up to $0.57a$ (with $a \sim 10^{-16}$
cm~\cite{Quigg1983}) is of electromagnetic in origin and the rest
of the mass in the form of the shell is of ordinary gravitational
mass.

\subsection{Case-III : $8 \pi p - E^2 = p_0$}
Following the previous case of constant charged density let us
consider here the {\it ansatz} for a constant pressure, say $p_0$
which is the central pressure of the fluid sphere and makes
balance between the attractive fluid pressure and repulsive
Coulombian force as an effective pressure ($p_{effective}$). In
reality the physical situation may not be so simple rather very
complicated one. However, we assume $8 \pi p - E^2 = p_0$ for
mathematical simplicity and would like to study an idealized
charged fluid model with conformal motion.

Now, from equation~(\ref{eqK2}), one can get an expression for
$\psi$ as
\begin{equation}
 \psi ^2  =  \frac{ C_3^2}{3}  (1 + p_0 r^2).
\end{equation}

\begin{figure}
\begin{center}
\vspace{0.5cm} \psfig{file=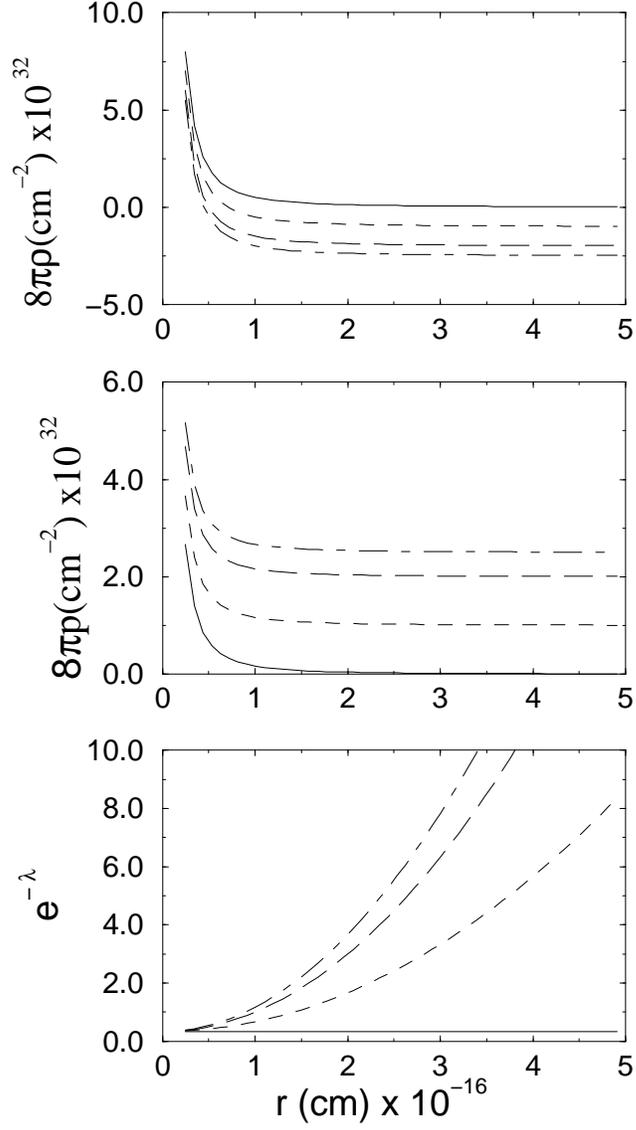,width=0.6\textwidth}
\caption{The upper, middle and lower panels represent plots for
$8\pi\rho$,  $8\pi p$ and $e^{-\lambda}$. The solid, dashed,
long-dashed and chain curves correspond to $p_0=0.0$, $1.0$, $2.0$
and $2.5$ cm$^{-2}$, respectively. } \label{fig:2}
\end{center}
\end{figure}

Hence all the other parameters can be obtained as
\begin{equation}
8\pi \rho  = -p_0 + \frac{1}{2r^2},
\end{equation}
\begin{equation}
8\pi p  =  p_0 + \frac{1}{6r^2},
\end{equation}
\begin{equation}
e^\nu  =C_2^2 r^2,
\end{equation}
\begin{equation}
e^{-\lambda}  = \frac{1}{3}  (1 + p_0 r^2),
\end{equation}
\begin{equation}
E^2  =  \frac{1}{6r^2},
\end{equation}
\begin{equation}
\sigma  = \frac {C_4} {4 \pi \sqrt{6} r}.
\end{equation}

The total gravitational mass $m(r=a)$ in this case can be
expressed as
\begin{equation}
\label{eqm3}
m(a) = \frac{1}{6}\left(\frac{5a}{2} - p_0a^3\right).
\end{equation}
We observe that $E^2$ and $\sigma$ with respect to their values of
case I (i.e. $p=-\rho$) are 1/3 and $1/\sqrt{3}$ times smaller,
respectively. The $e^{\nu}\propto r^2$ with a proportionality
constant $C_2^2$. For $E^2=8\pi\rho$, $p_0$ is zero. This implies
that $E^2=8\pi p=24\pi\rho$ too has $1/r^2$ behaviour and
$e^{-\lambda}$ with a value 1/3 is a constant. For the case
$E^2\neq 8\pi\rho$, equation~\ref{eqm3} applies an upper limit on
the choice of $p_0$ such that $p_0=8\pi p-E^2 < 5/2a^2$. The lower
limit being $0$ (as negative $p_0$ means $8\pi p <E^2$) we should
limit $p_0$ between $0$ and $5/2a^2$. It can, therefore, be easily
observed that for the condition $p_0=5/2a^2$ the above
gravitational mass becomes EMM whereas for $\rho$ the condition is
$p_0=3E^2$ and that for $p$ is $p_0=-E^2$. We plot all the
observable of this case in figure \ref{fig:2}.

\section{Concluding remarks}
We have analyzed the behaviour of a Lorentz's `extended electron'
within the framework of general theory of relativity and observed
that conformal motions do admit historically important
`electromagnetic mass' models. To do so we have studied three
special cases, viz., (1) $p = - \rho$, (2) $\sigma e^{\lambda/2} =
\sigma_0$ and (3) $8 \pi p - E^2 = p_0$. The results of the
present investigations are, in a nutshell, as follows.

{\it Case-I}: It has been observed that vanishing charge density
$\sigma$ makes $\psi_0=0=C_4$. This immediately implies
$E^2=8\pi\rho=-8\pi p=1/2r^2$ and $e^{\nu}=e^{-\lambda}=0$. Thus,
all the physical parameters, including the total gravitational
mass $m$, vanish due to vanishing charge. These results imply that
EMM model admitting conformal motion under the constraint
$C_4<\sqrt3/2q$.

{\it Case-II}: Here the expression for the total charge being
$q(a) = \sqrt{A} a^{3}$ we observe that for $\sigma_0=0$ the
constant $A(=16 \pi^2 \sigma_0^2 /9)$ becomes zero. This readily
implies that for the gravitational mass to vanish the condition to
be imposed here is $C_5=1/2a^2$ whereas for the energy density and
fluid pressure this is $C_5=\pm 1/6r^2$. Hence, we get a unique
condition for {\it EMM} model which can be expressed as
$r_{critical}=a/\sqrt3$. This suggests that not the mass of the
entire charged sphere rather only a radius up to $0.57a$ is of
electromagnetic in origin and the rest of the mass in the form of
the shell is of ordinary gravitational mass. In this connection we
would like to mention that in the framework of general theory of
relativity the electron-like spherically symmetric distribution of
matter must contain some negative mass
density~\cite{Cooperstock1989,Bonnor1989,Herrera1994,Ray2004b}.
Bonnor and Cooperstock~\cite{Bonnor1989} argue that the negativity
of the gravitational mass and hence negative energy density
($\rho<0$) for electron of radius $a \sim 10^{-16}$ is consistent
with the Reissner-Nordstr{\"o}m repulsion.

{\it Case-III}: We observe that for $E^2=8\pi\rho$, the constant
pressure term $p_0$ is zero. For the case $E^2\neq 8\pi\rho$ we
have an upper limit on the choice of $p_0$ such that $p_0=8\pi
p-E^2 < 5/2a^2$. The lower limit being $0$ we should limit $p_0$
between $0$ and $5/2a^2$. It can then be observed that for the
condition $p_0=5/2a^2$ the gravitational mass is of elctromagnetic
in origin whereas for $\rho$ the condition is $p_0=3E^2$ and that
for $p$ is $p_0=-E^2$.

Besides the above discussions we would also like to make here the
following comments which appear very much relevant in connection
to the present investigations:

(1) The equation of state in the form $p + \rho = 0$ is discussed
by Gliner~\cite{Gliner1966} in his study of the algebraic
properties of the energy-momentum tensor of ordinary matter
through the metric tensors and called it the $\rho$-vacuum state
of matter. It is also to be noted that the gravitational effect of
the zero-point energies of particles and electromagnetic fields
are real and measurable, as in the Casimir
Effect~\cite{Casimir1948}. According to Peebles and
Ratra~\cite{Peebles2003}, like all energies, this zero-point
energy has to contribute to the source term in Einstein's
gravitational field equation.

(2) In the Introduction we have mentioned that it is appropriate
to use inheritance symmetry to search for the natural relation
between geometry and matter. As a well known inheritance symmetry
we, therefore, exploit here the symmetry under conformal killing
vectors. The features with non-null CKV, as obtained in the
present work, show that electron has some probable inheritance
symmetry.

(3) In the present investigation we have employed the conformal
motion technique in connection to electron-like micro-particle
under the $4$-dimensional Einstein-Maxwell space-time. We,
therefore, feel that it may also be possible to extrapolate the
present investigation to the astrophysical bodies, specially quark
stars, admitting a one-parameter group of conformal motions. On
the other hand, the present investigation with $4$-dimensional
space-time can be extended to the higher dimensions to see the
features with conformal motions.

(4) It is argued by Gr{\o}n~\cite{Gron1986a,Gron1986b} that the
negative mass and the associated gravitational repulsion is due to
the strain of the vacuum because of vacuum polarization. He also
argues that if a vacuum has a vanishing energy, then its
gravitational mass will be negative and the observed expansion of
the universe may be explained as a result of repulsive
gravitation. It may also be pointed out that according to Ipser
and Sikivie~\cite{Ipser1984} domain walls are sources of repulsive
gravitation and a spherical domain wall will collapse. To overcome
this situation the charged ``bubbles" with negative mass keep the
wall static and hence in equilibrium.

\section*{Acknowledgments} Authors (SR and AAU) are
thankful to the authority of Inter-University Centre for Astronomy
and Astrophysics, Pune, India for providing them Associateship
programme under which a part of this work was carried out.

\end{document}